\pdfoutput=1
\documentclass[twoside,twocolumn]{article}

\usepackage{textcomp}
\usepackage{graphicx}
\usepackage[sc]{mathpazo} 
\usepackage[T1]{fontenc} 
\usepackage{microtype} 

\usepackage[english]{babel} 

\usepackage[hmarginratio=1:1,top=20mm,columnsep=20pt,lmargin=1.7cm,rmargin=1.7cm]{geometry} 
\usepackage{layout}
\usepackage[small,labelfont=bf,up,textfont=it,up]{caption} 
\usepackage{booktabs} 
\usepackage{enumitem} 
\setlist[itemize]{noitemsep} 

\usepackage{abstract} 

\usepackage{titlesec} 
\renewcommand\thesection{\arabic{section}} 
\renewcommand\thesubsection{\arabic{subsection}} 
\titleformat{\section}[block]{\normalsize\centering}{\thesection.}{0.2em}{} 
\titleformat{\subsection}[block]{\normalsize}{\thesubsection.}{1em}{} 

\usepackage{fancyhdr} 
\usepackage{titling} 

\usepackage{hyperref} 

\pretitle{\begin{center}\LARGE\bfseries} 
\posttitle{\end{center}} 
\title{Measuring the Nonlinear Refractive Index of Graphene using the Optical Kerr Effect Method} 
\author{%
\normalsize\textbf{Evdokia Dremetsika$^{1,*}$, Bruno Dlubak$^2$, Simon-Pierre Gorza$^1$, Charles Ciret$^1$,} \\[0.3ex] 
\normalsize\textbf{Marie-Blandine Martin$^3$, Stephan Hofmann$^3$, Pierre Seneor$^2$,  Daniel Dolfi$^2$,} \\[0.3ex] 
\normalsize\textbf{ Serge Massar$^1$, Philippe Emplit$^1$, Pascal Kockaert$^1$} \\[0.3ex] 
    \footnotesize\itshape
                               ${}^{1}$~OP\'ERA-photonique,
                               Universit\'e libre de Bruxelles,
                               CP\,194/5,
                               50, Av. F.\,D.~Roosevelt,
                               B-1050 Brussels (Belgium)\\%
                 \footnotesize\itshape
                               ${}^{2}$~Unit\'e Mixte de Physique, CNRS, Thales, Univ. Paris-Sud, Univ. Paris-Saclay, 
                               Avenue A. Fresnel 1,
                               91767 Palaiseau Cedex (France)\\%
		 \footnotesize\itshape
                                ${}^{3}$~Department of Engineering, 
                               University of Cambridge,
                               Cambridge CB3 0FA
                             (United Kingdom)\\
 \footnotesize \href{mailto: edremets@ulb.ac.be}{* edremets@ulb.ac.be} 
}
\date{} 
\renewcommand{\maketitlehookd}{
\begin{abstract}
By means of the ultrafast optical Kerr effect method coupled to optical heterodyne detection (OHD-OKE), we characterize the third order nonlinear response of graphene at telecom wavelength, and compare it to experimental values obtained by the Z-scan method on the same samples. From these measurements, we estimate a negative nonlinear refractive index for monolayer graphene, 
{\(n_2=-1.1\times10^{-13}\,\mathrm{m^2/W}\)}. This is in contradiction to previously reported values, which leads us to compare our experimental measurements obtained by the OHD-OKE and the Z-scan method with theoretical and experimental values found in the literature, and to discuss the discrepancies, taking into account parameters such as doping.  \copyright  2016 Optical Society of America.
\end{abstract}
}

\begin{document}

\maketitle

\section{Introduction}
\small
Since its isolation in 2004, graphene, a single sheet of carbon atoms in honeycomb lattice, has attracted the interest of many researchers for its unique properties and its potential use in many applications ~\cite{novoselov_roadmap_2012}.
Graphene is characterized by broadband optical characteristics, that are used more and more for photonics~\cite{bonaccorso_graphene_2010}. It was first used as a saturable absorber in mode-locked lasers and modulators~\cite{sun_graphene_2010,li_ultrafast_2014}, but is also a promising material for integrated photonics as it is ultrathin and compatible with CMOS technology. Graphene has been successfully tested for electro-optic applications~\cite{hu_broadband_2016}, and could even be used for all-optical signal processing on photonic integrated circuits, as theoretical predictions~\cite{mikhailov_nonlinear_2008,ooi_waveguide_2014,cheng_numerical_2015,semnani_nonlinear_2016,mikhailov_quantum_2016} and experimental results~\cite{hendry_coherent_2010, zhang_z-scan_2012, chen_nonlinear_2013,miao_broadband_2015} demonstrate that it has a high and broadband optical nonlinearity.

As all-optical signal processing relies usually on the nonlinear refractive index \(n_2\), the practical development of applications requires a good knowledge of both the sign and the magnitude of this coefficient, as well as its response time. However, while most experimental results reported so far demonstrate a high and broadband nonlinearity of graphene, they do not agree on the order of magnitude of \(n_2\). Experimental values were obtained with four-wave mixing~\cite{hendry_coherent_2010} and Z-scan~\cite{zhang_z-scan_2012, chen_nonlinear_2013,miao_broadband_2015}. From the two methods, only Z-scan is able distinguish between phase and amplitude effects and therefore to resolve the sign of \(n_2\). Even between the reported values of \(n_2\) measured with Z-scan, there is a difference of three orders of magnitude. 

The need for additional experimental works is not only justified by the discrepancies between reported experimental results, but also by the disagreements between existing theoretical works. In the first theoretical predictions about the third order nonlinear optical response of  graphene~\cite{mikhailov_nonlinear_2008,hendry_coherent_2010} the authors did not comment on the sign of \(n_2\). 
In the last few years, more detailed theoretical works have been published. In~\cite{semnani_nonlinear_2016, cheng_numerical_2015, ooi_waveguide_2014} it is suggested that the sign of the nonlinearity changes with the chemical potential. Chatzidimitriou \textit{et al.}~\cite{chatzidimitriou_rigorous_2015} performed a detailed comparison between some theoretical works to conclude that there is a considerable disagreement between them in terms of the magnitude and the sign of~\(n_2\).

In this Letter, we address the problem of the sign and the magnitude of the nonlinear refractive index of graphene from an experimental point of view. To this end, we implement the ultrafast optical Kerr effect method, coupled to optical heterodyne detection (OHD-OKE)~\cite{smith_optically-heterodyne-detected_2002} to characterize monolayer CVD (chemical vapor deposition) graphene on quartz, which is a substrate with very low \(n_2\). This method should eliminate some drawbacks of the Z-scan technique~\cite{falconieri_thermo-optical_1999,yang_distortions_2003}, as it is not sensitive to thermal nonlinearities and also is not intrinsically sensitive to inhomogeneities of the samples. Moreover, it provides the time evolution of the nonlinear response. Although OHD-OKE is a well-known technique for the measurement of the third-order optical nonlinearity, this is the first time to the best of our knowledge that it is applied to graphene. Therefore, we compare the results obtained with OHD-OKE, with results from the commonly-used Z-scan technique~\cite{sheik-bahae_sensitive_1990}. We had to perform our own measurement, as all Z-scan traces reported in the literature~\cite{zhang_z-scan_2012, chen_nonlinear_2013,miao_broadband_2015} were obtained from multilayer graphene, whereas, in our work we used monolayer graphene. Our experimental results show that the nonlinear refractive index of monolayer CVD graphene on quartz is negative, implying a self-defocusing nonlinearity.

\section{Methods}
The monolayer graphene film that we used in our experiments was grown by catalyzed chemical vapor deposition (CVD) on a \(\mathrm{Cu}\) layer using \(\mathrm{CH}_4\) precursor at \(\sim\)1000\textdegree{C}. A typical transfer step assisted by a poly(methyl methacrylate) resist film was carried to etch the \(\mathrm{Cu}\) in ammonium persulfate and deposit the \(\mathrm{cm^2}\) scale graphene layer on the quartz plate~\cite{kidambi_parameter_2012}. After the whole process, Raman spectroscopy measurements (at 532\,nm) showed ratios of the D peak  (resp. 2D peak) intensity to that of the G peak around 2\% (resp. higher than 2)~\cite{kidambi_parameter_2012}, demonstrating the high crystalline quality of the graphene monolayer~\cite{ferrari_interpretation_2000}. As described in~\cite{kidambi_parameter_2012}, the chemical doping was evaluated to be p-doped with a carrier concentration of a few~\(10^{12}\,\mathrm{cm^{-2}}\).

Finding its origins in ultrafast spectroscopy, the OHD-OKE method~\cite{mcmorrow_femtosecond_1988,kinoshita_low_1996, smith_optically-heterodyne-detected_2002} has been used for many years to study third order nonlinear optical phenomena and their dynamics in various materials. It is essentially a pump-probe technique which detects polarization changes in the probe caused by the pump-induced birefringence or dichroism of the material under study. The optical heterodyne detection~\cite{levenson_polarization_1979} is used to increase the signal to noise ratio, as well as to measure separately the nonlinear refractive index and the nonlinear absorption coefficient of the material. From these measurements, it is possible to calculate the third-order susceptibility of the material \(\chi^{(3)}\).
\begin{figure}[h]
\centering
\includegraphics[width=7cm]{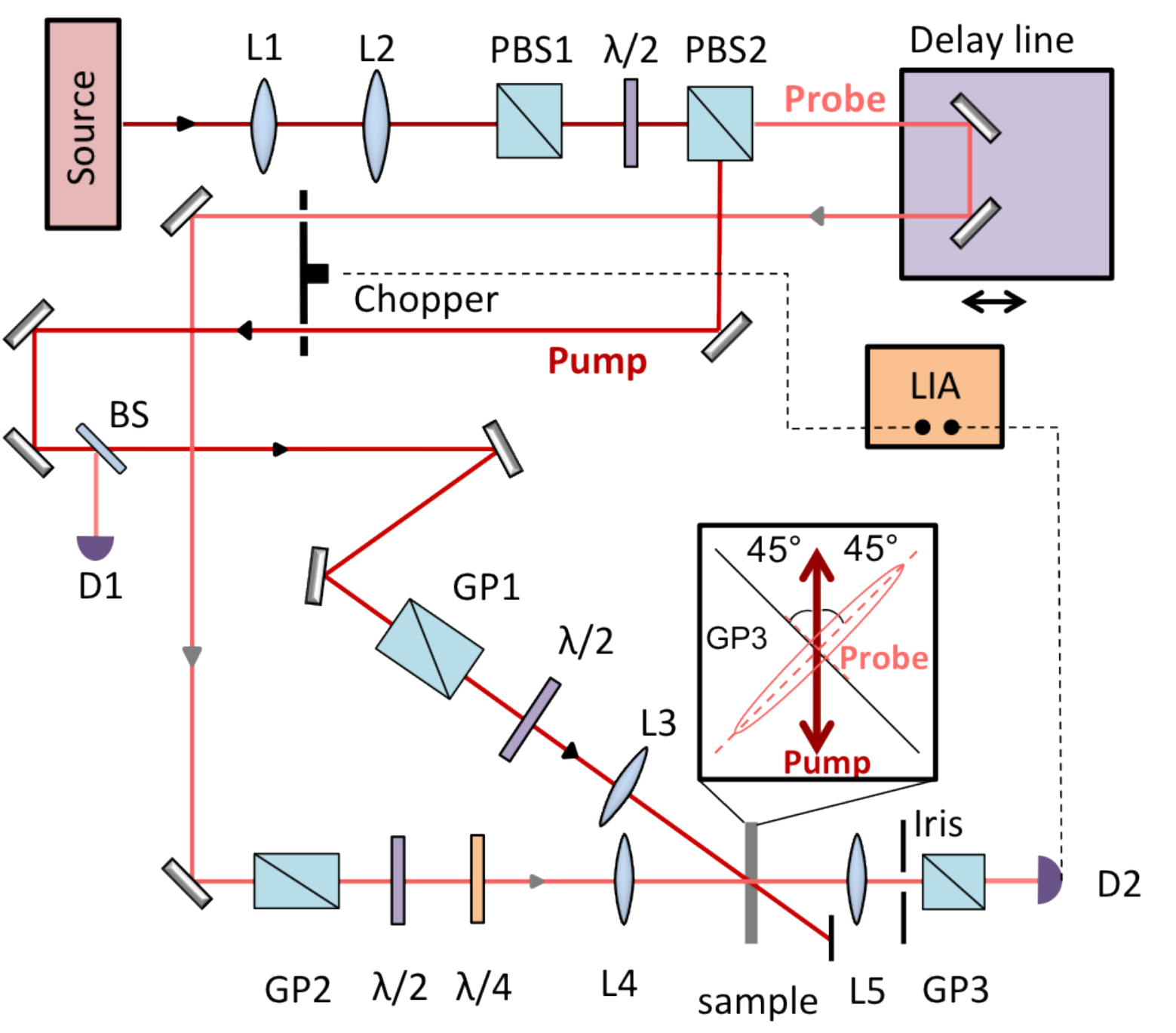}
\caption[Experimental setup for the OHD-OKE method.]
          {Experimental setup for the OHD-OKE method. The optical source delivers 180-fs pulses at 1600\,nm. Lenses \(L_1\) and \(L_2\) are used to adjust the size of the beam. Polarizing beam splitters PBS1 and PBS2 and the half-waveplate (\(\lambda/2\)), are used to control the power ratio between the pump and the probe. 
        The polarization of the pump and probe beams are tuned using the quarter- and half-wave plates (resp. \(\lambda/4\) and \(\lambda/2\)) and the Glan polarizers (GP1 and GP2) in order to obtain the polarization states depicted in the inset. The ratio between the long and the short axis of the probe polarization state is \(\tan\theta\).
        The chopper modulates the pump and the probe beams at \(f_5=5\,f_{w}\) and \(f_6=6\,f_{w}\) respectively, where \(f_{w}=41\,\mathrm{Hz}\) is the frequency of the wheel.
        }
\label{fig:setup}
\end{figure}

Our experimental setup is depicted in Fig.~\ref{fig:setup}. 
The excitation source is an optical parametric oscillator (OPO), pumped by a Ti:Sapphire laser, and delivering 180-fs pulses at a wavelength of 1600\,nm and a repetition rate of  82\,MHz. The signal is divided by means of a polarizing beam splitter in a strong pump and a weak probe with a ratio tuned from 10:1 to 20:1. A computer controlled motorized translation stage is used to vary the delay between the pump and the probe pulses.
As depicted in the inset of Fig.~\ref{fig:setup}, the polarization of the pump is set vertical, so that it lies in the graphene plane. The polarization of the probe beam lies also in this plane, but it is first set at 45\textdegree\ with respect to the pump polarization so that it is completely blocked by the output Glan polarizer (GP3) in the absence of nonlinear effect. High quality Glan polarizers (GP1, GP2) are used in order to define the polarizations of the two beams as well as possible. At a second stage, a slight rotation by an angle \(\theta/2\) of the half-wave plate on the probe beam leads to a small ellipticity of the probe polarization, while keeping its main axes at 45\textdegree\ with respect to the pump polarization. This small ellipticity enables the optical heterodyne detection. 
The pump and probe beams are spatially superimposed on the sample and focused down to a beam waist of \(w_3=20\,\mbox{\textmu{m}}\) and \(w_4=15\,\mbox{\textmu{m}}\) respectively, by means of the two lenses \(L_3\) (\(f_3=30\,\mathrm{mm}\)) and \(L_4\) (\(f_4=25\,\mathrm{mm}\)) respectively. The pump intensity is in the range of 2 to \(5\times10^{12}\,\mathrm{W/m^2}\), which is far below the damage threshold of graphene~\cite{currie_quantifying_2011}. The pump beam is blocked after the sample with an iris and the probe beam passes through the output Glan polarizer (GP3) before it is detected on the detector D2.

We implemented the optically heterodyned detection by setting \(\theta\neq0\), and by chopping the pump and probe beams at different frequencies, as proposed in~\cite{alavi_optically_1990}. A dual-phase lock-in amplifier processed the output of detector \(D_2\) (Ge-based), in order to extract the nonlinear signal at the sum of the modulation frequencies. In this way, we avoid measuring the scattering of the pump or other background signals.

\section{Results}

When the measurement is performed at \(\theta=0\), the induced birefringence modifies the probe polarization at high pump intensity, leading to a non-zero field \(E_{NL}\) after GP3~\cite{smith_optically-heterodyne-detected_2002}. The intensity of the signal at zero delay after lock-in detection at the sum of the modulation frequencies is proportional to \(S_{homo}=\left\vert{\frac{2\pi}{\lambda}\cdot{}(n_2+\mathrm{i}\kappa_2)\cdot{}L\cdot{I_{pump}}}\right\vert^2\cdot{I_{probe}}\), where \(I_{pump}\) and \(I_{probe}\) are respectively the pump and the probe intensities, \(L\) is the effective interaction length between the pump and the probe, and \(\kappa_2\) accounts for the nonlinear absorption. As can be seen, \(S_{homo}\) does not provide information on the sign of \(n_2\). Optically heterodyned detection occurs when \(\theta\neq0\). In this case, the polarization component on the short axis of the polarization ellipsis of the probe beam plays the role of what is generally called the local oscillator field \(E_{LO}\). This local oscillator interferes with the nonlinear signal \(E_{NL}\), so that a new component \(2\,E_{LO}\cdot{}E_{NL}\) appears in the detected signal, which is proportional to \(E_{LO}\), and therefore to~\(\theta\). The full signal after lock-in detection is in a first approximation \(S_{OHD-OKE}=S_{homo}+\theta\cdot{}S_{hetero}\), where \(S_{hetero}=(\frac{2\pi}{\lambda}\cdot{}n_2\cdot{}L\cdot{I_{pump}})\cdot{I_{probe}}\)~\cite{smith_optically-heterodyne-detected_2002}. We used \(\theta\) around \(4\mbox{\textdegree}\) to obtain a good signal-to-noise ratio. 
In practice, we measure \(S_{OHD-OKE}\) at different delays between the pump and the probe signal, which allows to temporally resolve the nonlinear response of the material.
\thispagestyle{empty}

\begin{figure}[h]
\centering
\includegraphics[width=0.49\linewidth]{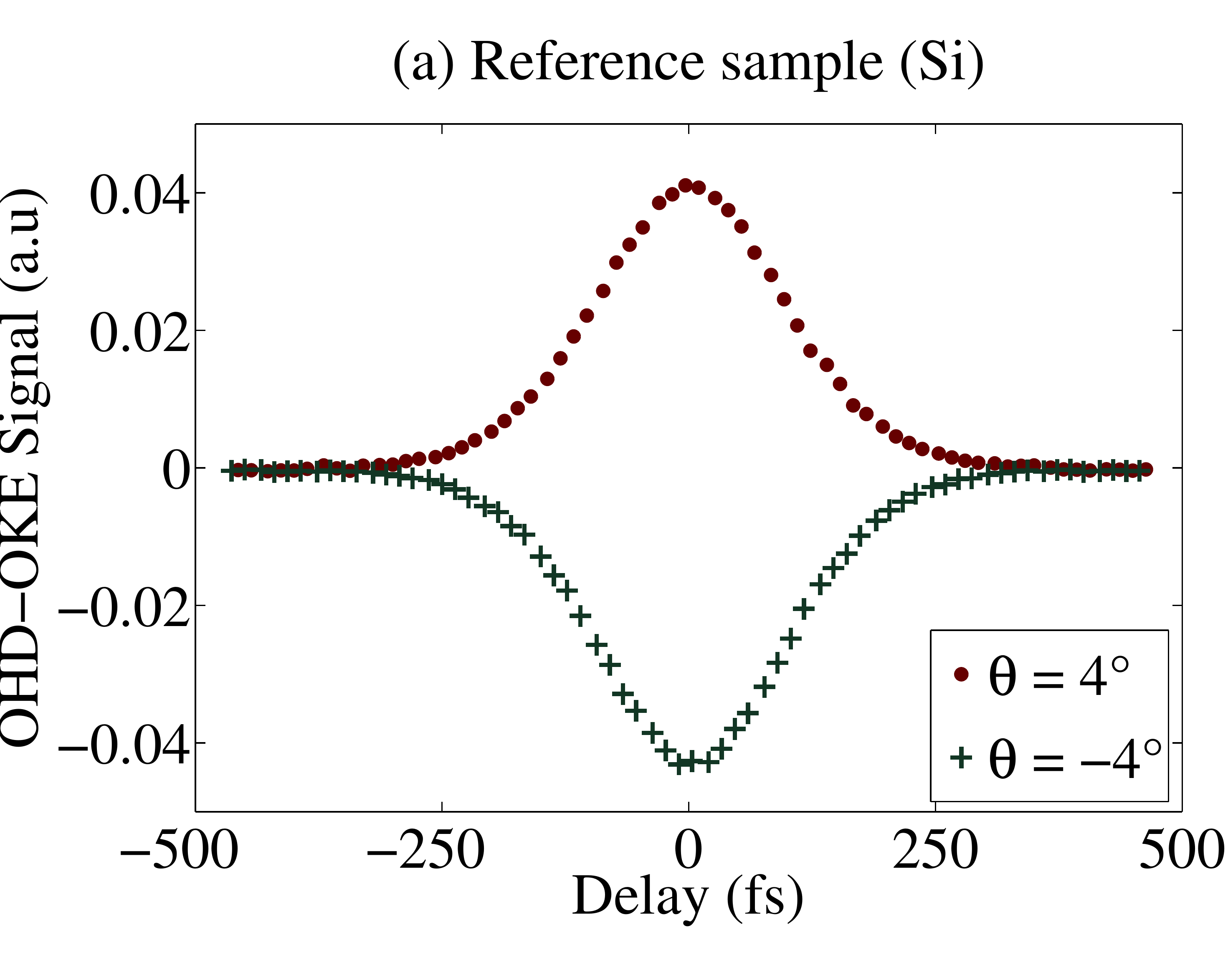}\hfill
\includegraphics[width=0.49\linewidth]{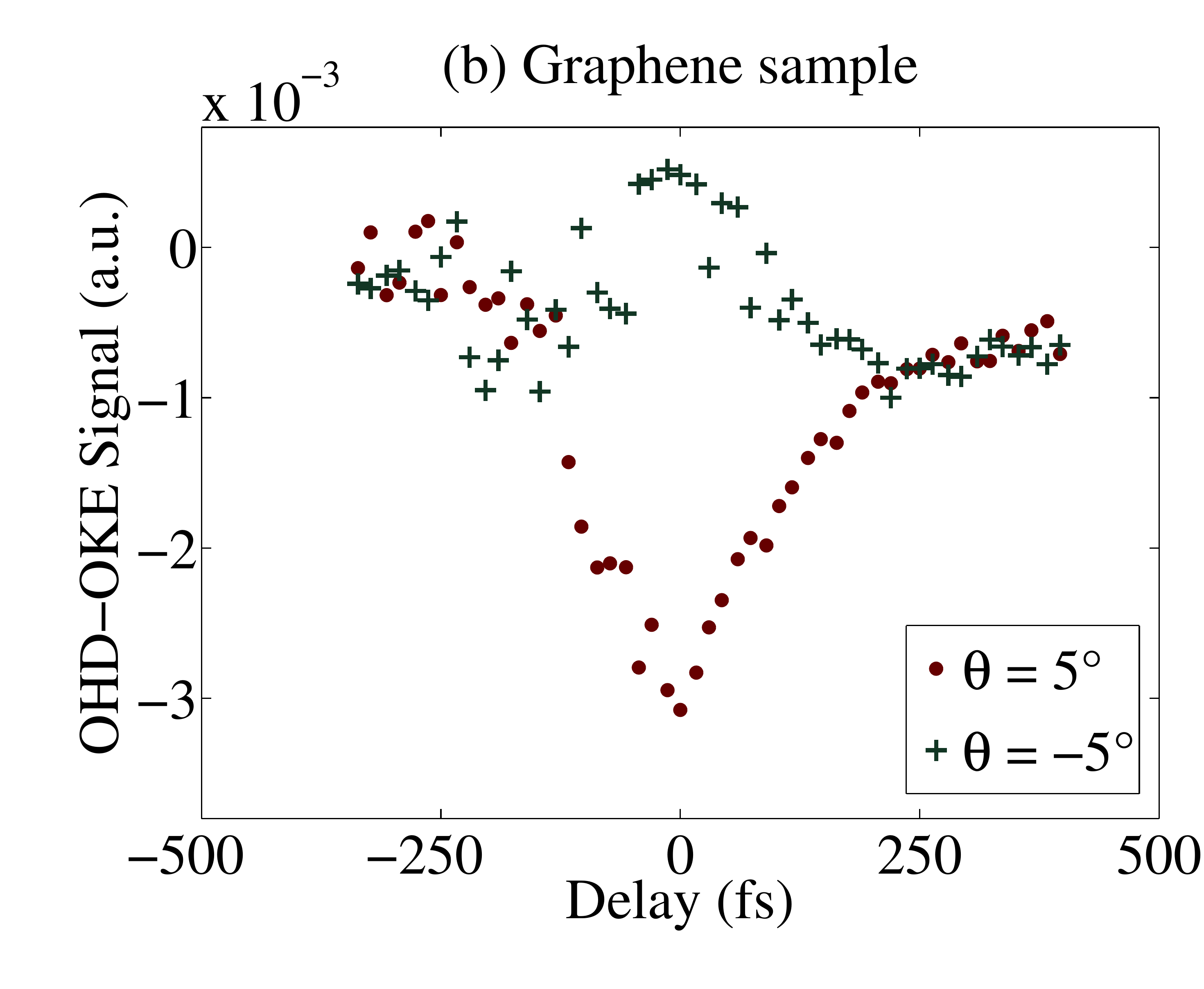}
\includegraphics[width=0.49\linewidth]{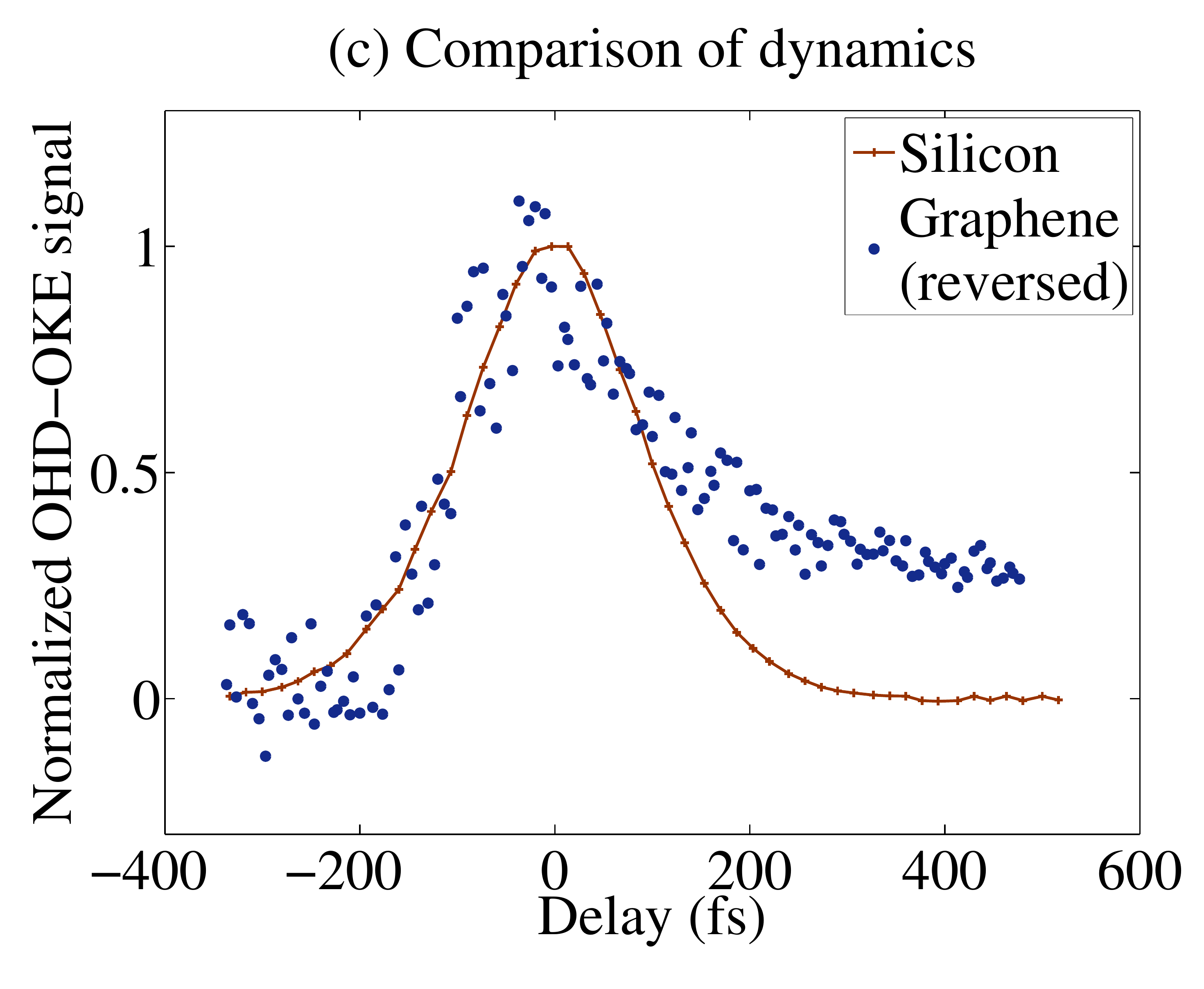}
\caption[Experimental results from OHD-OKE]
        {Experimental results from OHD-OKE: (a)~OHD-OKE signal of reference sample (Si). (b)~OHD-OKE signal of monolayer CVD graphene. The vertical scales are in arbitrary units. We can see that the curves in (a) and~(b) corresponding to angles \(\theta\) present reversed signs. 
        (c)~Normalized value of OHD-OKE signal of monolayer CVD graphene and silicon at \(\theta=4\mbox{\textdegree}\), showing the delayed response of graphene. To allow for a better comparison, the sign of the graphene signal has been reversed. }
\label{fig:OKEresults}
\end{figure}

We first tested our setup on a reference sample made of 1\,mm thick pure silicon. Due to the limited overlap between the pump and the probe beams, the effective interaction length was \(L\approx100\,\mbox{\textmu{m}}\). We found a value of the nonlinear refractive index of silicon \(n_2^{Si}=5.5\times10^{-18}\,\mathrm{m^2/W}\) at a wavelength of 1600\,nm, which is close to previously reported values~\cite{dinu_third-order_2003}. The OHD-OKE signals at positive and negative heterodyne angles \(\theta=\pm4\mbox{\textdegree}\) are shown in Fig.~\ref{fig:OKEresults}(a). As the FWHM of the OHD-OKE signal with respect to the delay between the pump and the probe corresponds to the autocorrelation of 180-fs laser pulses, our results  indicate an ultrafast nonlinearity with electronic origin. 

The graphene sample was then used to record the OHD-OKE signal in Fig.~\ref{fig:OKEresults}(b). (Note that the asymmetry is due to nonlinear absorption, which is not studied in this Letter.) We observe in Fig.~\ref{fig:OKEresults}(c)  a slower response at the picosecond scale which is typical for graphene~\cite{dawlaty_measurement_2008,breusing_ultrafast_2011}, and is due to interband transitions.

\begin{figure}[h]
\centering
\includegraphics[width=4.2cm]{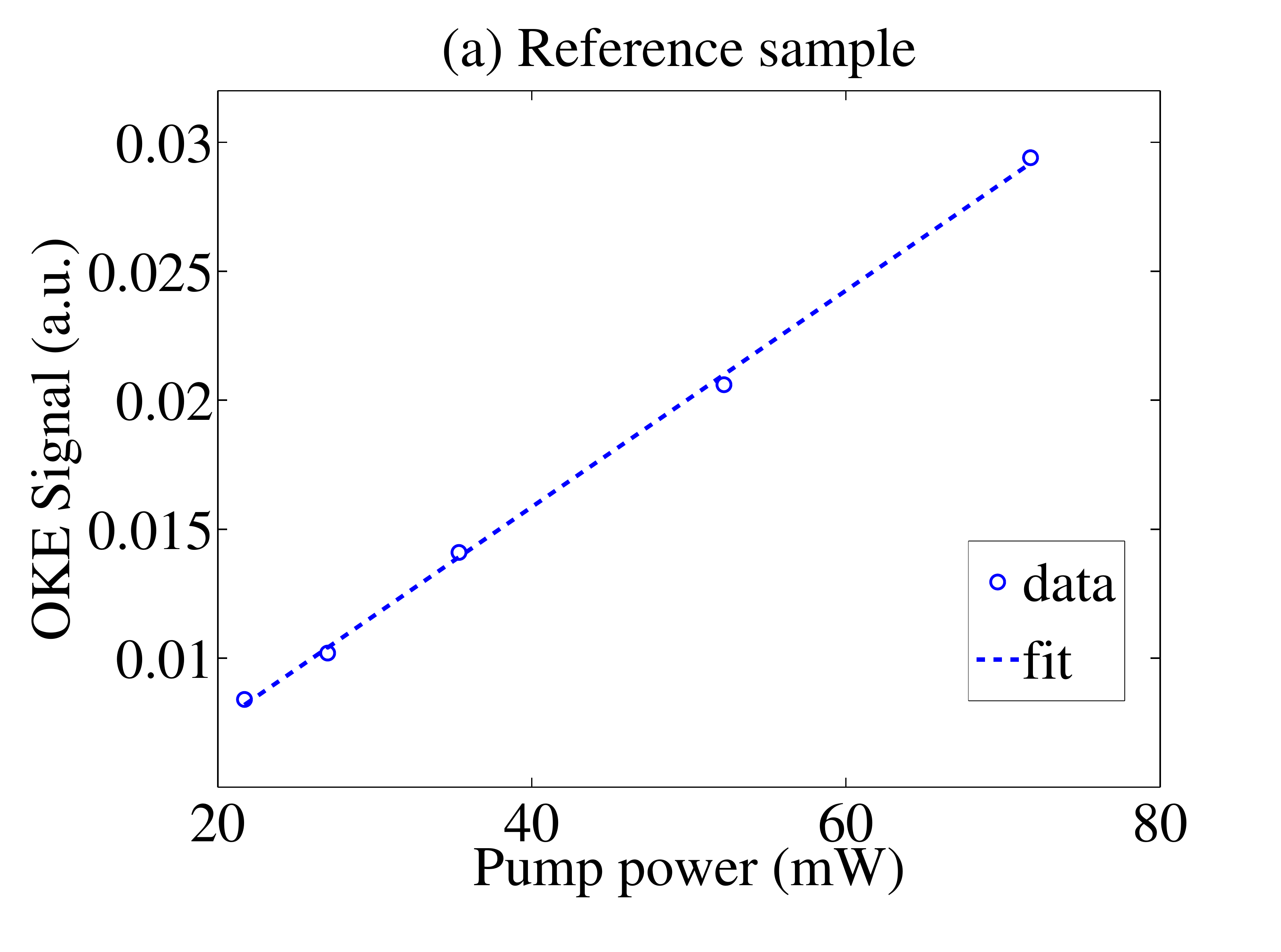}
\includegraphics[width=4.2cm]{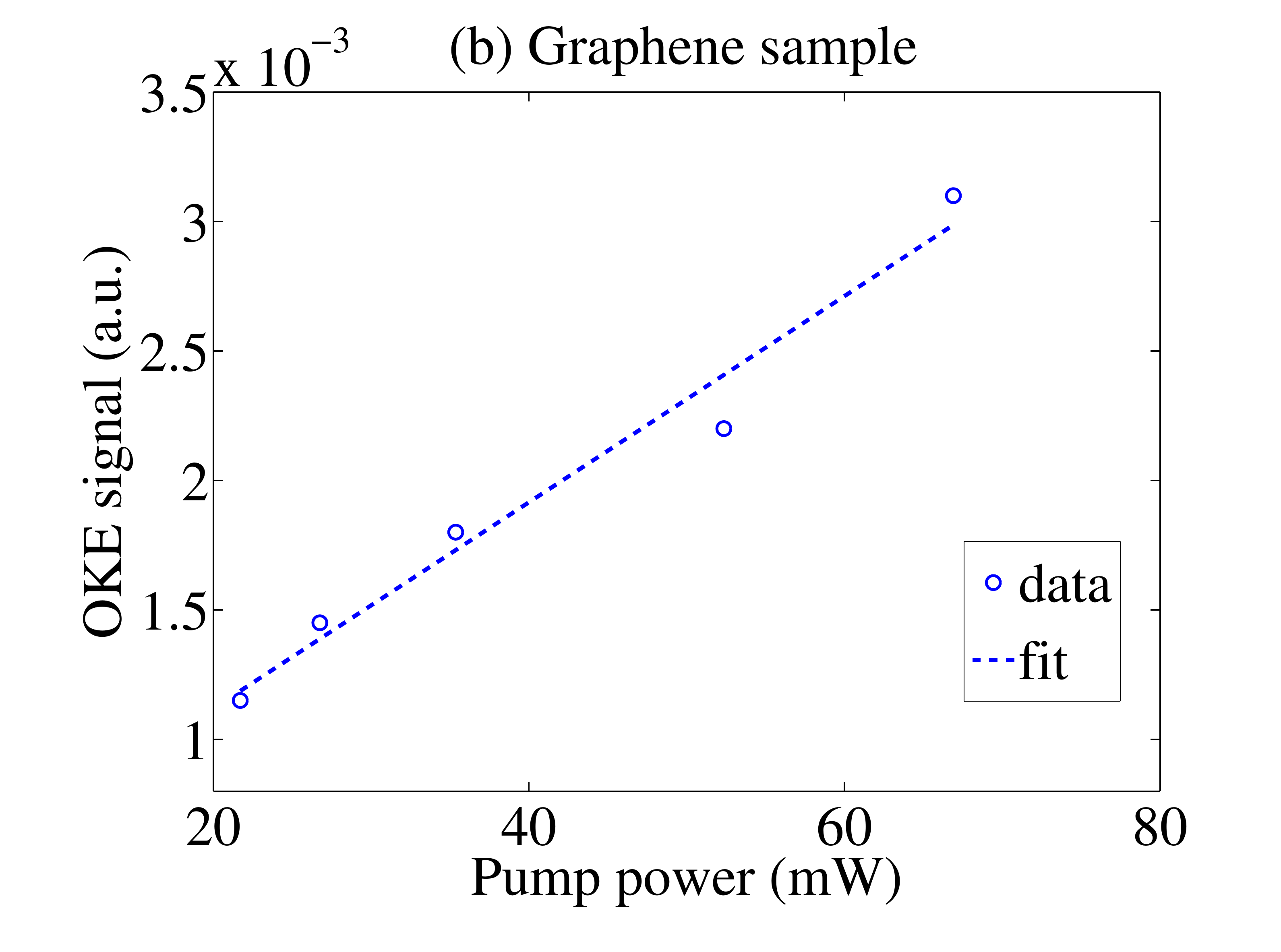}
\caption[]{Dependence of OHD-OKE signal at zero delay with the pump power.}
\label{fig:SignalPump}
\end{figure}

In order to further test the reliability of our setup, we performed measurements of the OHD-OKE signal at zero delay while varying the pump power separately for the reference and the graphene samples. The results shown in Fig.~\ref{fig:SignalPump} demonstrate that the retrieved signal is linearly dependent on the pump power, which is expected if it results from the expression of \(S_{hetero}\). We also confirmed that the OHD-OKE signal at zero delay is minimized when the pump polarization is set parallel or orthogonal to the probe polarization~\cite{smith_optically-heterodyne-detected_2002}. The OHD-OKE signal in a region of the substrate without graphene, was measured to be below the detection limit of our setup showing that the contribution of the substrate to the results presented in Fig.~\ref{fig:OKEresults} is negligible.

We estimate the nonlinear refractive index of graphene by comparison with the reference sample using the following formula
\begin{equation}
n_2^s = \frac{S_{hetero}^s}{S_{hetero}^r}\frac{L^r}{L^s}\,n_2^r,
\label{eq:eqokeref}
\end{equation}
where \(L^r\) is the effective length of the reference sample (with losses included) limited by the confocal length, \(L^s\) is the effective length of the graphene sample, \(n_2^r\) the measured nonlinear refractive index of the reference (Si) and \(S_{hetero}^r\), \(S_{hetero}^s\) are the heterodyne parts of the signal for the reference and the graphene sample respectively (at zero delay). 
For each sample, we estimate \(S_{hetero}\) by taking the difference between OHD-OKE signals obtained with opposite values of~\(\theta\). Repeated measurements at different locations on the sample provide a negative nonlinear refractive index for graphene, with a mean value of \(n_2=-1.07\times10^{-13}\,\mathrm{m^2/W}\), and a standard deviation \(\sigma_{n_2}=0.26\times10^{-13}\,\mathrm{m^2/W}\). The main systematic error that could affect our measurement is the estimate of the ratio \(L^r/L^s\), which could modify the magnitude of \(n_2\), but not its sign. This systematic error is due to the difficulty to measure the confocal length of the reference sample and to define the thickness of one graphene layer.

\thispagestyle{empty}

In order to verify that the nonlinear refractive index of our monolayer CVD graphene samples is negative, we performed Z-scan measurements on the same samples. The Z-scan method is a single-beam technique~\cite{sheik-bahae_sensitive_1990} that provides the magnitude and the sign of the nonlinear refractive index by monitoring the transmission of the central part of the beam through a closed aperture in the far field, while scanning a thin sample through the focal point of a focused laser beam. By using an open aperture, the same scan provides a direct measurement of the nonlinear absorption, which is beyond the scope of this Letter. As the beam quality is of high importance for the Z-scan method, especially in our case where the sample is ultrathin, we used a fiber laser to perform these measurements. The experimental setup is the same as in~\cite{zhang_z-scan_2012}. The optical source was a Pritel fiber laser amplified by a Pritel EDFA, delivering 3.8-ps pulses at 1550\,nm wavelength with a repetition rate of 10 MHz.
The peak-on-axis intensity was \(5\times10^{12}\,\mathrm{W/m^2}\), which is below the damage threshold of graphene~\cite{currie_quantifying_2011}. 
\begin{figure}[h]
\centering
\includegraphics[width=7cm]{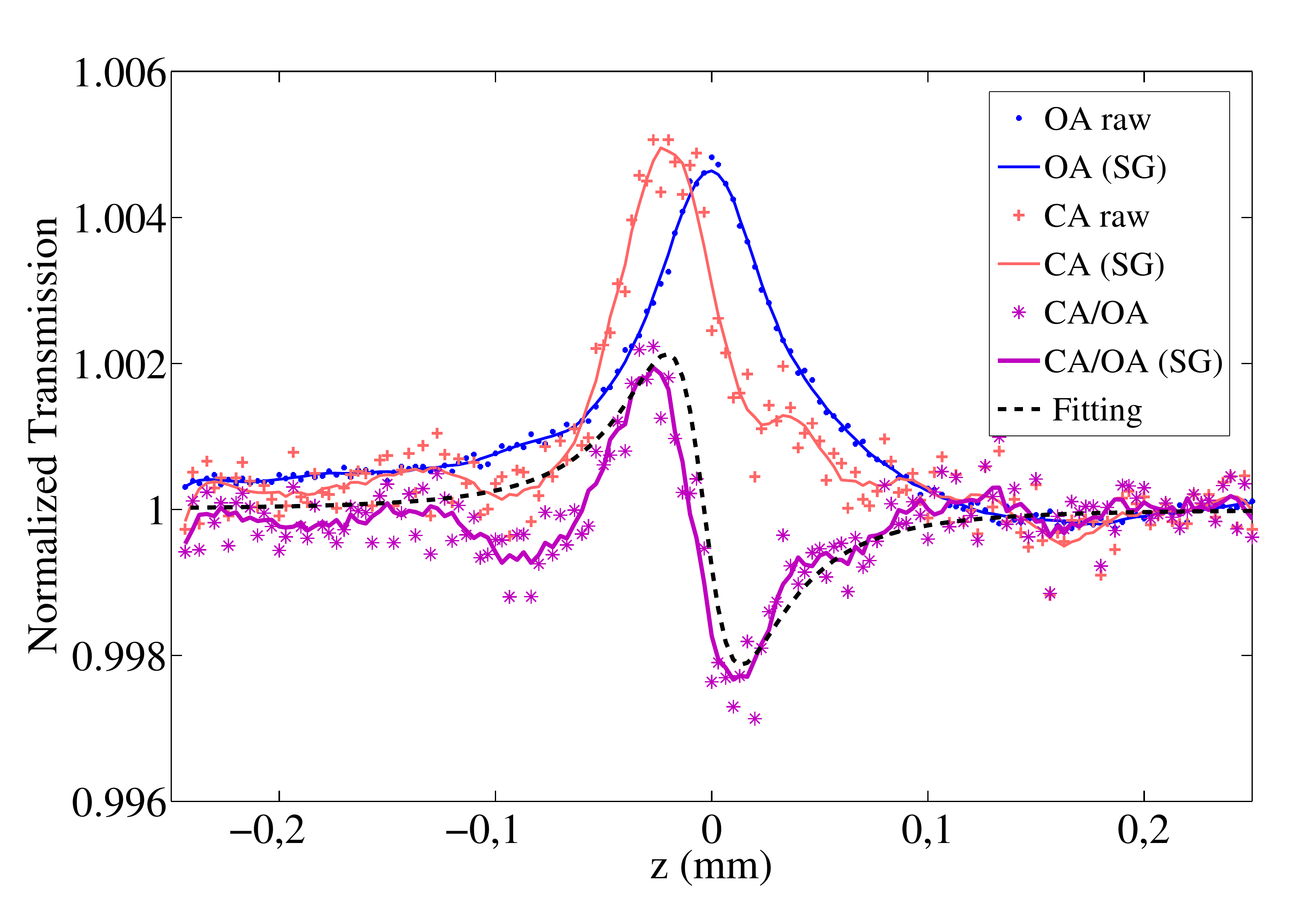}
\caption{Z-scan trace of monolayer graphene resulting from the division of the closed aperture (CA) data by the open aperture (OA) data. A Savitzky-Golay (SG) filtering of the data is also provided for better visibility. The fitting is performed on the SG-filtered data of the Z-scan trace.}
\label{fig:Zscan}
\end{figure}
A typical trace for monolayer graphene is depicted in Fig.~\ref{fig:Zscan}. The closed-aperture trace divided by the open-aperture trace results in a typical peak-valley Z-scan trace, which indicates a self-defocusing medium~\cite{sheik-bahae_sensitive_1990}. For the fitting we used the formula from~\cite{sheik-bahae_sensitive_1990}: \(T(x) = 1+(4x\Delta\Phi)/[(1+x^2)(9+x^2)]\), where  \(x=z/z_R\) is the normalized position to the Rayleigh length \(z_R\) and \(\Delta\Phi=(2\pi/\lambda)n_2LI_0\), with \(L\) and \(I_0\) being the graphene thickness and peak-on-axis intensity respectively. 
We verified that thermal effects were negligible by performing the same measurement with an optical chopper that modified the mean power and the repetition rate without changing the peak power. We also checked that the response of the substrate was negligible. The high quality of the samples ensured that the measurement was not affected by inhomogeneities of graphene, which was confirmed by the repeatability of the results at different locations on the sample. The estimated value of the nonlinear refractive index from these measurements is \(n_2=-2\times10^{-12}\,\mathrm{m^2/W}\). 
\thispagestyle{empty}

\section{Discussion}
We have deduced a negative nonlinear refractive index for graphene at telecommunication wavelengths from both OHD-OKE and Z-scan experiments, in contradiction to the previously-reported Z-scan experimental works~\cite{zhang_z-scan_2012, chen_nonlinear_2013,miao_broadband_2015}. 
However, in Fig.3~\cite{zhang_z-scan_2012} the Z-scan trace is typical for a negative nonlinearity. Therefore, either the figure is correct and there is an error in the sign of \(n_2\) that could be due to a typographic error of sign in the fitting formula published in~\cite{sheik-bahae_sensitive_1990}, or the sign of \(n_2\) is correct and the discrepancy with our measurements could be due to the chemical doping of graphene samples. According to theoretical works~\cite{cheng_numerical_2015,ooi_waveguide_2014} reviewed in~\cite{chatzidimitriou_rigorous_2015}, the sign and the magnitude of the nonlinearity of graphene depend on the chemical potential, which can be modified by applying a voltage or by chemical doping. This conclusion is supported by the latest results reported in~\cite{semnani_nonlinear_2016,mikhailov_quantum_2016}. The authors of~\cite{berciaud_probing_2009} report that CVD graphene transferred on quartz is slightly p-doped. Also, Roberts \textit{et al.}~\cite{roberts_response_2011} suggest that after exposure to laser pulses and oxygen for long periods, graphene samples can become highly p-doped. Nevertheless, they conclude that working with intensities below \(10^{14}\,\mathrm{W/m^2}\) should avoid this effect.

\thispagestyle{empty}
As for the magnitude of \(n_2\), the value obtained with the Z-scan method is one order of magnitude higher than the one deduced from OHD-OKE measurements. 
We attribute this to the fact that the OKE signal at zero delay represents only the fast electronic nonlinearity, while the Z-scan with 3.8-ps pulses includes the nonlinear processes due to interband transitions as well. 
This could explain also the fact that our result from OHD-OKE is of the same order of magnitude as the one obtained in the four-wave mixing experiment reported in~\cite{hendry_coherent_2010}, whereas our Z-scan value is closer in magnitude to results from~\cite{zhang_z-scan_2012}. It is probably for the same reason that Z-scan results with femtosecond pulses~\cite{chen_nonlinear_2013} provide lower values for \(n_2\) than our experiment with picosecond pulses. Such differences could be expected according to~\cite{nikolakakos_broadband_2004}.

\section{Conclusion}
To summarize, we used the ultrafast OHD-OKE method to estimate the nonlinear refractive index of monolayer graphene on quartz as \(n_2=-1.1\times10^{-13}\,\mathrm{m^2/W}\). The OHD-OKE method, unlike Z-scan, has the advantage of being insensitive to thermal nonlinearities and sample inhomogeneities. However, both methods indicate that graphene has a self-defocusing nonlinearity, which contradicts previously reported results. We have discussed possible explanations for this disagreement. We hope that this Letter will help to clarify the contradictions that appear between different theoretical works, as it seems that some works did not focus their attention on the sign of the nonlinear refractive index. We also believe that there is a need for additional theoretical and experimental investigation on the relation between the third order optical nonlinearity and the chemical potential of graphene.

\textbf{Funding.} Belgian Science Policy Office (BELSPO) under grant IAP7-35; EU FP7 Work Program (Graphene Flagship No. 604391, Core1 No. 696656); E. D. is funded by the Fund for Research Training in Industry and Agriculture (FRIA, Belgium).  
\thispagestyle{empty}

\end{document}